\documentclass[10pt,a4paper]{article}

\usepackage[breaklinks]{hyperref}
\usepackage[T1]{fontenc} 
\usepackage[latin1]{inputenc}

\usepackage{aeguill}
\usepackage{graphicx}
\usepackage{amsfonts}
\usepackage{amsmath}
\usepackage{amssymb}
\usepackage{amssymb,stmaryrd}
\usepackage{multirow}
\usepackage{garamond}
\usepackage[urw-garamond]{mathdesign}
\usepackage[sort&compress]{natbib}
\usepackage[]{url,hyperref}

\usepackage[ruled,linesnumbered]{algorithm2e}

\newenvironment{keywords}{
       \list{}{\advance\topsep by0.35cm\relax\small
       \leftmargin=1cm
       \labelwidth=0.35cm
       \listparindent=0.35cm
       \itemindent\listparindent
       \rightmargin\leftmargin}\item[\hskip\labelsep
                                     \bfseries Keywords:]}
     {\endlist}

\begin{document}

\title{IRM4MLS:\\the influence reaction model for multi-level simulation}

\author{
\\~\\~\\
Gildas Morvan$^{1,2}$~~~Alexandre Veremme$^{1,3}$~~~Daniel Dupont$^{1,3}$\\~\\~\\
$^{1}$Univ Lille Nord de France, F-59000 Lille, France\\
$^{2}$UArtois, LGI2A, F-62400, Béthune, France\\
$^{3}$HEI, F-59046, Lille, France\\~\\~\\
\url{http://www.lgi2a.univ-artois.fr/~morvan/}\\\url{gildas.morvan@univ-artois.fr}\\~\\~\\\includegraphics[width=3cm]{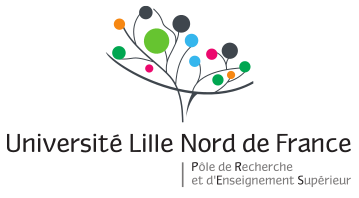}~~\includegraphics[width=3cm]{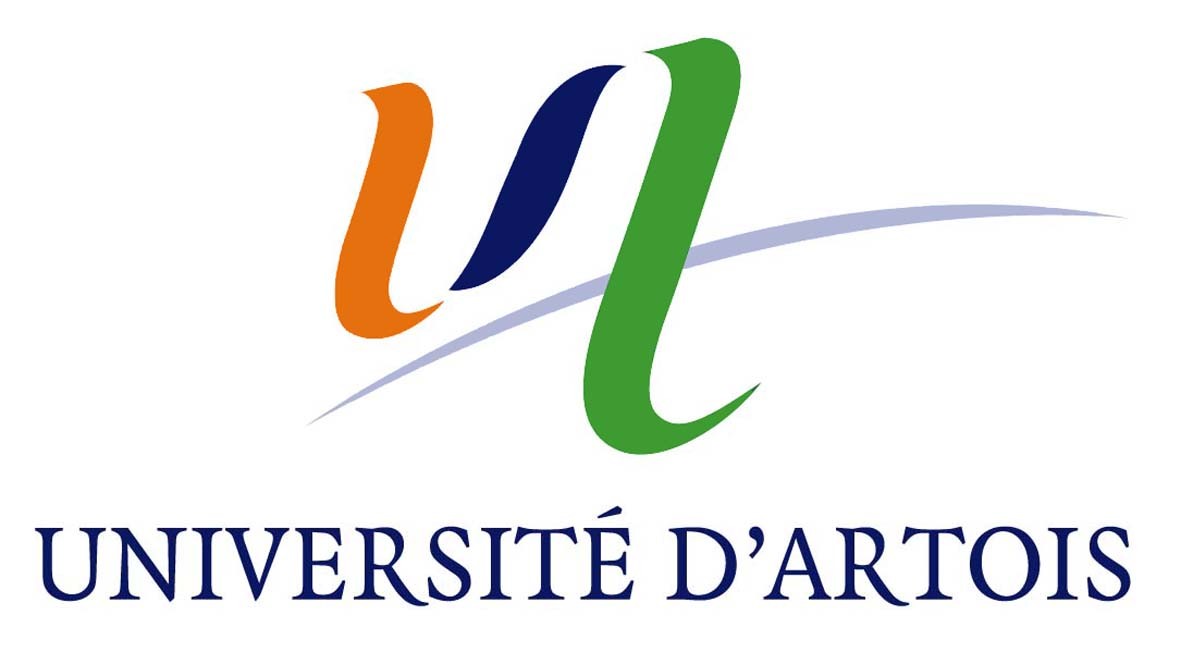}~~\includegraphics[width=2.5cm]{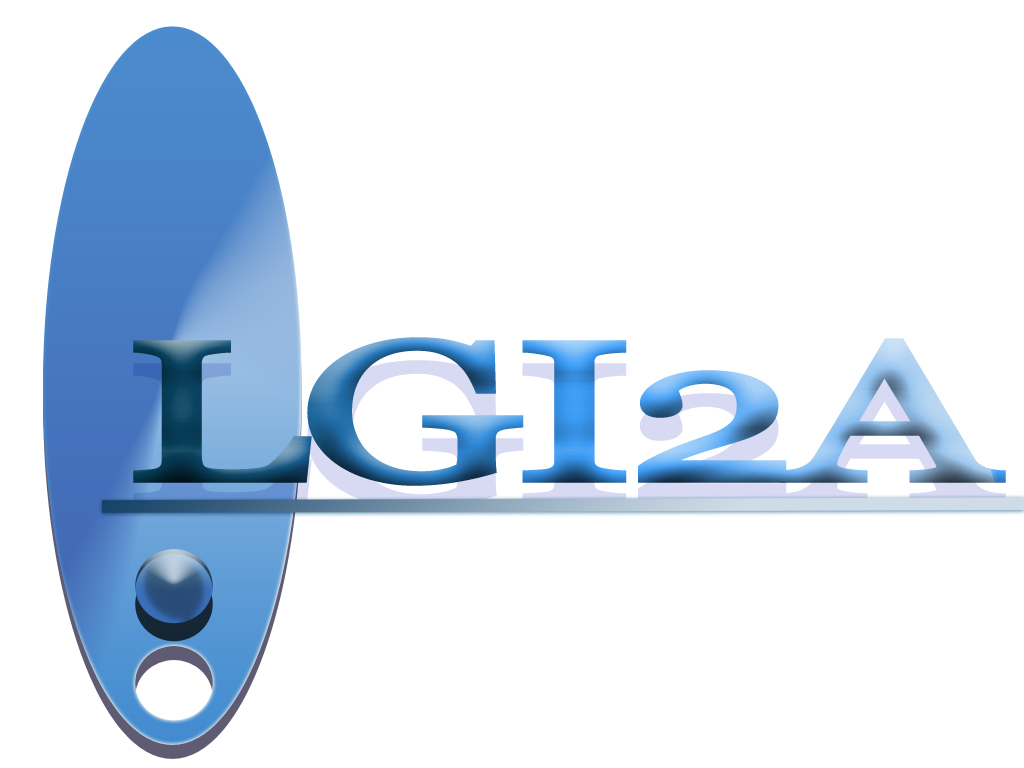}~~\includegraphics[width=4.2cm]{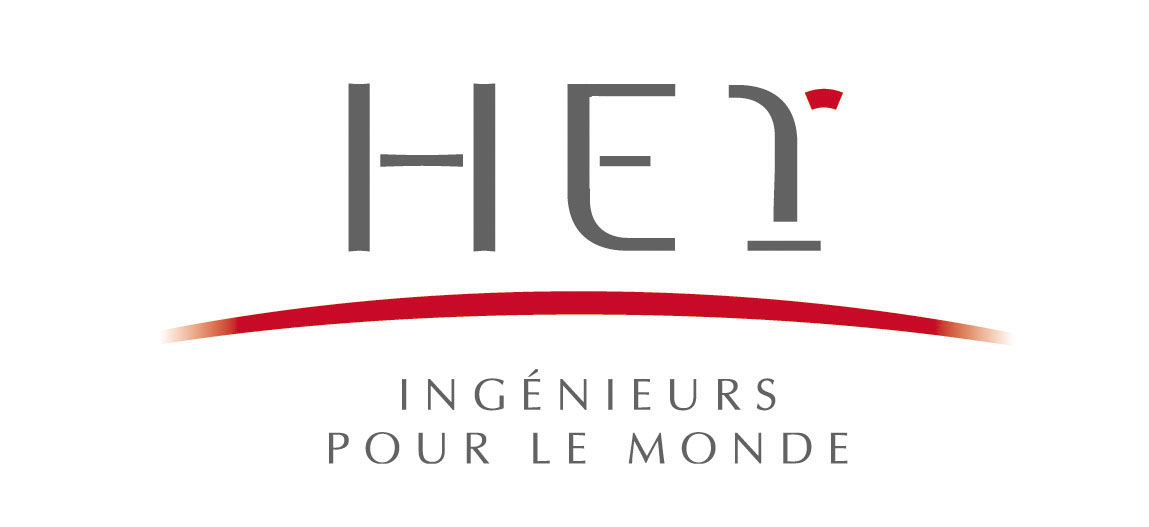}\\~\\~\\~\\~\\
}

\date{}

\garamond

\maketitle

\newpage

\begin{abstract}
In this paper, a meta-model called IRM4MLS, that aims to be a generic ground to specify and execute multi-level agent-based models is presented. It relies on the influence/reaction principle and more specifically on IRM4S~\citep{Michel:2007,Michel:2007a}. Simulation models for IRM4MLS are defined. The capabilities and possible extensions of the meta-model are discussed. 
\end{abstract}


\section{Introduction}

The term "multi-level modeling" refers to the modeling of a system considered at various levels of organization. \textit{E.g.},  a biological system can be considered at different levels: 
\begin{center}
... $\rightarrow$ molecule $\rightarrow$  cell $\rightarrow$ tissue  $\rightarrow$ organ  $\rightarrow$ ... , 
\end{center}
that basically correspond to the segmentation of biological research into specialized communities:
\begin{center}
... $\rightarrow$ molecular biology $\rightarrow$ cell biology $\rightarrow$ histology $\rightarrow$ physiology  $\rightarrow$ ... .
\end{center}

Each research area has developed its own ontologies and models to describe the same reality observed at different levels. However, this reductionist approach fails when considering complex systems. \textit{E.g.}, it has been shown that living systems are co-produced by processes at different levels of organization~\citep{Maturana:1980}. Therefore, an explanatory model of such systems should consider the interactions between levels. Agent-based modeling (ABM) is a paradigm of choice to study complex systems. But, while it seems more interesting to integrate knowledge from the different levels studied and their interactions in a single model, ABM often remains a pure bottom-up approach~\citep{Drogoul:2003a}.  

Thus, recently\footnote{It has to be noted that the eleven year old model RIVAGE pioneered the field of ML-ABM~\citep{Servat:1998}.} various research projects have aimed at developing multi-level agent-based models (ML-ABM) in various fields such as histology, ethology or sociology~\citep{An:2008,Gil-Quijano:2008,Lepagnot:2009,Morvan:2008,Morvan:2009a,Pumain:2009,Zhang:2009}. A good analysis of some of these models, and the motivations of these works can be found in~\citet{Gil-Quijano:2009}. 

Various issues should be addressed when developing a ML-ABM. For instance one major problem is the automatic detection of emergent phenomena that could influence other levels~\citep{Chen:2009,David:2009,Prevost:2009}. Another important problem is the temporal and spatial mappings of model levels and thus the scheduling of the simulations~\citep{Hoekstra:2007}. More exhaustive presentations of these issues can be found in~\citet{Morvan:2009a,Gil-Quijano:2009}.

In the models found in literature, these issues  have been addressed according to the specificity of the problem. Indeed, they are based on \textit{ad-hoc} meta-models and the transferability of ideas from one to another seems difficult.  

In this paper, a meta-model that aims to be a generic ground to specify and execute ML-ABM is presented.  It is based on  IRM4S  (an Influence Reaction Model for Simulation) proposed in~\citet{Michel:2007,Michel:2007a}, itself based on  IRM (Influences and Reaction model) originally presented in~\citet{Ferber:1996}. IRM4S is described in section~\ref{IRM4S} and its multi-level extension, called IRM4MLS  (Influence Reaction Model for Multi-level Simulation), in section~\ref{MLM}. Section~\ref{simulation} introduces two simulation models for IRM4MLS. The first one is very simple and similar to IRM4S but supposes that all levels have the same temporal dynamics while the second one has a more general scope but relies on temporal constraints and thus, is more complicated and time consuming.

\section{The IRM4S meta-model}
\label{IRM4S}

IRM was developed to address issues raised by the classical vision of action in Artificial Intelligence as \textit{the transformation of a global state}: simultaneous actions cannot be easily handled, the result of an action depends on the agent that performs it but not on other actions and the autonomy of agents is not respected~\citep{Ferber:1996}.  

While IRM addresses these issues, its complexity makes it difficult to implement. IRM4S is an adaptation of IRM, dedicated to simulation, that clarifies some ambiguous points. It is described in the following.

Let $\delta(t) \in \Delta$ be the dynamic state of the system at time $t$:
\begin{equation}
   \delta(t) = < \sigma(t), \gamma(t) >, 
\end{equation}
where $\sigma(t) \in \Sigma$ is the set of environmental properties and $\gamma(t) \in \Gamma$ the set of influences, representing system dynamics. The state of an agent $a \in A$ is characterized by:
\begin{itemize}
\item necessary, its physical state $\phi_a \in \Phi_a$ with $\Phi_a \in \Sigma$ (\textit{e.g.}, its position),  
\item possibly, its internal state $s_a \in S_a$ (\textit{e.g.}, its beliefs).
\end{itemize}
Thus, IRM4S distinguishes between the mind and the body of the agents. 

The evolution of the system from $t$ to $t + dt$ is a two-step process:
\begin{enumerate}
\item agents and environment produce a set of influences\footnote{the sets of producible influence sets and influences produced at $t$ are denoted respectively $\Gamma'$ and $\gamma'(t)$ to point out that the latter is temporary and will be used to compute the dynamic state of the system at $t+dt$.} $\gamma'(t) \in \Gamma'$:
\begin{equation}
   \gamma'(t) = Influence(\delta(t)),
\end{equation}
\item the reaction to influences produces the new dynamic state of the system:
 \begin{equation}
  \delta(t+dt) = Reaction(\sigma(t),\gamma'(t)).
\end{equation}
\end{enumerate}
 As~\cite{Michel:2007a} notes, "the influences [produced by an agent] do not directly change the environment, but rather represent the desire of an agent to see it changed in some way". Thus, $Reaction$ computes the consequences of agent desires and environment dynamics.

An agent $a \in A$  produces influences through a function $Behavior_a: \Delta \mapsto \Gamma'$. This function is decomposed into three functions executed sequentially:
\begin{equation}
	p_a(t) = Perception_a(\delta(t)),
\end{equation}
\begin{equation}
	s_a(t+dt) = Memorization_a(p_a(t), s_a(t)),
\end{equation}
\begin{equation}
	\gamma'_a(t) = Decision_a(s_a(t+dt)).
\end{equation}

The environment produces influences through a function $Natural_\omega: \Delta \mapsto \Gamma'$:
\begin{equation}
	\gamma'_\omega(t) = Natural_\omega(\delta(t)).
\end{equation}

Then the set of influences produced in the system at $t$ is:
\begin{equation}
\gamma'(t) = \{\gamma(t) \cup \gamma'_\omega(t) \cup \bigcup_{a \in A} \gamma'_a(t) \}.
\end{equation}

After those influences have been produced, the new dynamic state of the system is computed by a  function $Reaction: \Sigma \times \Gamma' \mapsto \Delta$ such as: 
\begin{equation}
	\label{reaction}
	\delta(t+dt) = Reaction(\sigma(t), \gamma'(t)).
\end{equation}

Strategies for computing $Reaction$ can be found in~\cite{Michel:2007a}.

\section{The influence reaction model for multi-level simulation (IRM4MLS)}
\label{MLM}

\subsection{Specification of the levels and their interactions} 

A multi-level model is defined by a set of levels $L$ and a specification of the relations between levels. Two kinds of relations are specified in IRM4MLS: an influence relation (agents in a level $l$ are able to produce influences in a level $l' \neq l$) and a perception relation (agents in a level $l$ are able to perceive the dynamic state of a level $l' \neq l$), represented by directed graphs denoted respectively $<L, E_I>$ and $<L, E_P>$, where $E_{I}$  and $E_{P}$ are two sets of edges, \textit{i.e.}, ordered pairs of elements of $L$. Influence and perception relations in a level are systematic and thus not specified in $E_I$ and $E_P$ (cf. eq.~\ref{I-} and \ref{I+}). 

\textit{E.g.},$\forall l,l' \in L^2$, if  $E_{P} = \{ll'\}$ then the agents of $l$ are able to perceive the dynamic states of  $l$ and  $l'$ while the agents of $l'$ are able to perceive the dynamic state of  $l'$. 

The perception relation represents the capability, for agents in a level, to be "conscious" of other levels, \textit{e.g.}, human beings having knowledge in sociology are conscious of the social structures they are involved in. Thus, in a pure reactive agent simulation, $E_P = \emptyset$. $E_P$ represents what agents are able to be conscious of, not what they actually are: this is handled by a perception function, proper to each agent. 

The in and out neighborhood in  $<L, E_I>$ (respectively $<L, E_P>$)  are denoted $N_{I}^-$ and $N_{I}^+$ (resp. $N_{P}^-$ and $N_{P}^+$) and are defined as follows:
\begin{equation}
\forall l \in L, N_{I}^-(l) \mbox{ (resp. } N_{P}^-(l)\mbox{) } =  \{ l \} \cup \{l' \in L: l'l \in E_{I} \mbox{ (resp. } E_{P}\mbox{)} \},
\label{I-} 
\end{equation}
\begin{equation}
\forall l \in L, N_{I}^+(l) \mbox{ (resp. } N_{P}^-(l)\mbox{) } =  \{ l \} \cup \{l' \in L: ll' \in E_{I} \mbox{ (resp. } E_{P}\mbox{)} \},
\label{I+} 
\end{equation}

\textit{E.g.}, $\forall l,l' \in L^2$ if $l' \in N_I^+(l)$ then the environment and the agents of $l$ are able to produce influences in the level $l'$; conversely we have $l \in N_I^-(l')$, \textit{i.e.}, $l'$ is influenced by $l$.

 \subsection{Agent population and environments} 

The set of agents in the system at time $t$ is denoted $A(t)$. $\forall l \in L$, the set of agents belonging to $l$ at $t$ is denoted $A_l(t) \subseteq A(t)$. An agent belongs to a level iff a subset of its physical state $\phi_a$ belongs to the state of the level:
 \begin{equation}
\label{indicatorfunction}
\forall a \in A(t), \forall l \in L, a \in A_l(t) \text{~iff~} \exists \phi^l_a(t) \subseteq \phi_a(t) | \phi^l_a(t) \subseteq \sigma^l(t).
\end{equation}
Thus, an agent belongs to zero, one, or more levels. An environment can also belong to different levels.

 %A multi-level agent-based model is then defined as a tuple $<L, E_I, E_P>$. 
 
 \subsection{Influence production} 
 
 The dynamic state of a level $l \in L$ at time $t$, denoted  $\delta^l(t) \in \Delta^l$, is a tuple  $< \sigma^l(t), \gamma^l(t) >$, where $\sigma^l(t)  \in \Sigma^l$ and  $ \gamma^l(t) \in  \Gamma^l$ are the sets of environmental properties and influences of $l$. 
 
The influence production step of IRM4S is modified to take into account the influence and perception relations between levels. Thus,  the $Behavior_a^l$ function of an agent $a \in A_l$  is defined as:  
\begin{equation}
Behavior_a^l:  \prod_{l_P \in N_P^+(l)}  \Delta^{l_P}  \mapsto \prod_{l_I \in N_I^+(l)} \Gamma^{l_I}{'}.
\end{equation}

This function is described as a composition of functions. As two types of agents are considered (\textit{tropistic} agents, \textit{i.e.}, without memory and \textit{hysteretic} agents, \textit{i.e.}, with memory\footnote{While the tropistic/hysteretic distinction is made in IRM, it does not appear clearly  in IRM4S. However, in a multi-level context, it is important if multi-level agents are considered.}), two types of behavior functions are defined~\cite{Ferber:1999}.

An hysteretic agent $ha$ in a level $l$ acts according to its internal state. Thus, its behavior function is defined as:
\begin{equation}
Behavior_{ha}^l = Decision_{ha}^l \circ Memorization_{ha} \circ Perception_{ha}^l, 
\end{equation}
with
\begin{equation}
	\label{perception}
	Perception_{ha}^l: \prod_{l_P \in N_P^+(l)}  \Delta^{l_P}  \mapsto   \prod_{l_P \in N_P^+(l)} P_{ha}^{l_P},
\end{equation}
\begin{equation}
	Memorization_{ha}: \prod_{l \in L | ha \in A_l} \prod_{l_P \in N_P^+(l)} P_{ha}^{l_P} \times S_{ha} \mapsto S_{ha}, 
\end{equation}
\begin{equation}
	Decision_{ha}^l: S_{ha} \mapsto \prod_{l_I \in N_I^+(l)} \Gamma^{l_I}{'}.
\end{equation}
There is no memorization function specific to a level. Like in other multi-agent system meta-models ---\textit{e.g.}, MASQ \citep{Stratulat:2009}---, we consider that  an agent can have multiple bodies but only one mind (\textit{i.e.}, one internal state). Moreover, the coherence of the internal state of the agents would have been difficult to maintain with several memorization functions.

A tropistic agent $ta$ in a level $l$ acts according to its percepts:
\begin{equation}
Behavior_{ta}^l = Decision_{ta}^l  \circ Perception_{ta}^l, 
\end{equation}
with $Perception_{ta}^l$ following the definition of eq.~\ref{perception} and
\begin{equation}
	Decision_{ta}^l: \prod_{l_P \in N_P^+(l)} P_{ta}^{l_P} \mapsto \prod_{l_I \in N_I^+(l)} \Gamma^{l_I}{'}.
\end{equation}

The environment $\omega$ of a level $l$ produces influences through a function: 
\begin{equation}
	Natural_\omega^l: \Delta^{l} \mapsto \prod_{l_I \in N_I^+(l)} \Gamma^{l_I}{'}.
\end{equation}

 \subsection{Reaction to influences} 

Once influences have been produced, interactions between levels do not matter anymore. Thus,  the reaction function defined in IRM4S can be re-used: 
 \begin{equation}
	Reaction^l: \Sigma^l \times \Gamma^l{}' \mapsto \Delta^l,
\end{equation}
where $Reaction^l$ is the reaction function proper to each level. 

\section{Simulation of IRM4MLS models}
\label{simulation}

In this section, two simulation models for IRM4MLS are proposed. The first one (section~\ref{simplesimulation}) is directly based on  IRM4S. It supposes that all levels have the same temporal dynamics. The second one (section~\ref{complexsimulation}) has a more general scope but is also more complicated and time consuming.
These models are compatible with the different classical time evolution methods (event-to-event or fixed time step) used in multi-agent simulation.  In the following, $t_0$ and $T$ denote the first and last simulation times.

\subsection{A simple simulation model}
\label{simplesimulation}

In this section, a model with single temporal dynamics is introduced. As there is no synchronization  issue, it is very similar to the model of IRM4S. Eq.~\ref{temporalmodel1} to~\ref{temporalmodeln}  describe this simple temporal model. $\mbox{HA}(t)$ and $\mbox{TA}(t)$ denote respectively the sets of hysteretic and tropistic agents in the system. 

First, behavior sub-functions are executed for each agent:
 \begin{equation}
	\label{temporalmodel1}
	\forall l \in L, p_a(t) = < Perception_a^l(<\delta^{l_P}(t): l_P \in N_P^+(l) >) : a \in A_l(t) >, 
\end{equation}
\begin{equation}
	\forall a \in \mbox{HA}(t), s_a(t+dt) = Memorization_a(p_a(t)),
\end{equation}
\begin{equation} 
	\forall l \in L, \forall a \in \mbox{HA}_l(t), <\gamma_a^{l_I}{'}(t): l_I \in N_I^+(l) >= Decision_a^l(s_a(t+dt)),
\end{equation}
\begin{equation}
	\forall l \in L, \forall a \in \mbox{TA}_l(t), <\gamma_a^{l_I}{'}(t): l_I \in N_I^+(l) > = Decision_a^l(p_a(t)).
\end{equation}

Then, environmental influences are produced:
\begin{equation}
	\forall l \in L, <\gamma_{\omega}^{l_I}(t): l_I \in N_I^+(l) > = Natural_\omega^l(\delta^l(t)).
\end{equation}
 
The set of temporary influences in a level $l \in L$ at $t$ is defined as:
\begin{equation}
\label{gammap}
\gamma^l{'}(t) = \{\gamma^l(t)   \bigcup_{l_I \in N_I^-(l)} \gamma^{l_I}_\omega{'}(t) \bigcup_{a \in A_{l_I}} \gamma_a^{l_I}{'}(t) \}.
\end{equation}

Finally, the new state of the system can be computed:
 \begin{equation}
	\label{temporalmodeln}
	\forall l \in L, \delta^l(t+dt) = Reaction^l(\sigma^l(t),  \gamma^l{}'(t)).
\end{equation}

Algorithm~\ref{simplealgo} summarizes this simulation model. 
\begin{algorithm}[h]
%\SetLine
\KwIn{$<L, E_I, E_P >, A(t_0), \delta(t_0)$}
\KwOut{$\delta(T)$}
$t = t_0$\;
\While{$t \leq T$}{
   \ForEach{$a \in A(t)$}{
      $p_a(t) = <Perception_a^l(<\delta^{l_P}(t): l_P \in N_P^+(l) >) : a \in A_l >$\;
      \If{$a \in \mbox{HA(t)}$}{$s_a(t+dt) = Memorization_a(p_a(t))$\;}
   }
   \ForEach{$l \in L$}{
      $<\gamma_{\omega}^{l_I}{'}(t): l_I \in N_I^+(l) > = Natural_\omega^l(\delta^l(t))$\;
      \ForEach{$a \in \mbox{HA}_l(t)$}{
             $<\gamma_a^{l_I}{'}(t): l_I \in N_I^+(l) > = Decision_a^l(s_a(t+dt))$\;
         }
        \ForEach{$a \in \mbox{TA}_l(t)$}{
            $<\gamma_a^{l_I}{'}(t): l_I \in N_I^+(l) > =  Decision_a^l(p_a(t))$\;
        }
   }
   \ForEach{$l \in L$}{
      $\gamma^l{'}(t) = \{\gamma^l(t)   \bigcup_{l_I \in N_I^-(l)} \gamma^{l_I}_\omega{'}(t) \bigcup_{a \in A_{l_I}} \gamma_a^{l_I}{'}(t) \}$\;
      $\delta^l(t+dt) = Reaction^l(\sigma^l(t),  \gamma^l{}'(t))$ \;
   }
  $t = t+dt$\;
}
\caption{simple simulation model of IRM4MLS}
\label{simplealgo}
\end{algorithm}

 \subsection{A simulation model with level-dependent temporal dynamics}
\label{complexsimulation}

In this section, a simulation model with level-dependent temporal dynamics is introduced. In the following, $t^l$ and $t^l + dt^l$ denote respectively the current and next simulation times of a level $l \in L$. Moreover $t = < t^l : l \in L >$ and $t +dt = < t^l + dt^l : l \in L >$ denote respectively the sets of current and next simulation times for all levels. It is mandatory to introduce rules that constraint perceptions, influence production and reaction computation.  These rules  rely primarily on the \textit{causality principle}:
\begin{itemize}
\item an agent cannot perceive the future, \textit{i.e.}, 
 \begin{equation}
	\label{causalityP}
	\forall l \in L, l_P \in N_P^+(l)  \mbox{ is perceptible from } l \mbox{ if } t^l \geq t^{l_P},
\end{equation}
\item an agent or an environment cannot influence the past, \textit{i.e.}, 
 \begin{equation}
	\label{causalityI}
	\forall l \in L, l_I \in N_I^+(l)  \mbox{ can be influenced by } l \mbox{ if } t^l \leq t^{l_I}.
\end{equation}
\end{itemize}

However, the causality principle is not sufficient to ensure a good scheduling. A \textit{coherence principle} should also guide the conception of the simulation model:
\begin{itemize}
\item an agent can only perceive the latest available dynamic states, \textit{i.e.}, 
 \begin{equation}
	\label{coherenceP}
	\forall l \in L, l_P \in N_P^+(l)  \mbox{ is perceptible from } l \mbox{ if } t^l < t^{l_P} + dt^{l_P},
\end{equation}
\item as a hysteretic agent can belong to more than one level, its internal state must be computed for the next simulation time at which it is considered, \textit{i.e.}, 
 \begin{equation}
	\label{coherenceS}
	\forall l \in L, s_a(t_a + dt_a) = Memorization_a(p_a(t^l)),
\end{equation}
such as
\begin{equation}
	\begin{array}{c}
		t_a + dt_a = t^l + dt^l | \forall t^l{}' + dt^l{}',  t^l + dt^l \geq t^l{}' + dt^l{}'\\
		\Rightarrow t^l + dt^l =t^l{}' + dt^l{}' \wedge a \in A_l,
	\end{array}
\end{equation}
\item an agent or an environment can influence a level according to its latest state,  \textit{i.e.}, 
 \begin{equation}
	\label{coherenceI}
	\forall l \in L, l_I \in N_I^+(l)  \mbox{ can be influenced by } l \mbox{ if } t^l +dt^l > t^{l_I},
\end{equation}
\item reaction must be computed for the next simulation time, \textit{i.e.}, 
 \begin{equation}
	\label{coherenceR}
	\forall l \in L,  Reaction^l  \mbox{ is computed if }  t^l +dt^l  \in min(t + dt).
\end{equation}
\end{itemize}

Moreover, a \textit{utility principle} should also be applied:
\begin{itemize}
\item perceptions should be computed at once, \textit{i.e.}, 
 \begin{equation}
	\label{utilityP}
	\begin{array}{c}
		\forall l \in L, \forall a \in A_l, Perception_a^l  \text{ is computed}\\
		\text{if } \forall l_P \in N_P^+(l), t^l \geq t^{l_P}.
	\end{array}
\end{equation}
\item as well as influences, \textit{i.e.}, 
 \begin{equation}
	\label{utilityI}
	\begin{array}{c}
		\forall l \in L, Natural_\omega^l \mbox{ and } \forall a \in A_l, Decision_a^l  \text{ are computed}\\
		\text{if } \forall l_I \in N_I^+(l), t^l \leq t^{l_I} \vee t^l + dt^l <  t^{l_I} + dt^{l_I}.
	\end{array}
\end{equation}
\end{itemize}
 
It is easy to show that the rule defined in eq.~\ref{utilityP} subsums  the rule defined in eq.~\ref{causalityP}. Moreover, the rule defined in eq.~\ref{coherenceR} implies the rule defined in eq.~\ref{coherenceP}. 

According to eq.~\ref{utilityI}, influences are not necessarily produced at each time from a level $l$ to a level $l_I \in  N_I^+(l)$. Thus, a function $c_I$, defines influence production from the rules defined by the eq.~\ref{coherenceI} and~\ref{utilityP}:
\begin{equation}
	\forall l, \in L, \forall l_I \in N_I^+(l), c_I(l,l_I) = \left\{
		\begin{array}{clc}
			\gamma^{l_I}{}'(t^{l_I}) &\mbox{if}&t^l \leq  t^{l_I} \wedge  t^l +dt^l > t^{l_I}\\ 
			\emptyset&\mbox{else.}&
		\end{array}\right.
\end{equation}

The simulation model can then be defined as follows. First, if the condition defined in the eq.~\ref{utilityP} is respected, agents construct their percepts and consecutively hysteretic agents compute their next internal state:
\\~\\
$\forall a \in A(t),$ 
	\begin{align}
		p_a(t^l)& = < Perception_a^l(< \delta^{l_P}(t^{l_P}): l_P \in N_P^+(l) >) : l \in L_P>,\\
		s_a(t_a + dt_a)&= Memorization_a(p_a(t^l)) \mbox{ if } a \in \mbox{HA}(t),
	\end{align}
with $L_P = \{ l \in L : a \in A_l(t) \wedge \forall l_P \in N_P^+(l),  t^l \geq t^{l_P}\}$.

Then, if the condition defined in eq.~\ref{utilityI} is respected, agents and environments produce influences:
\\~\\
	$\forall l \in L_I,$
	\begin{align}	
		<c_I(l,l_I) : l_I \in N_I^+(l) >&=Natural_\omega^l(\delta^l(t^l)),\\
		\forall a \in \mbox{HA}_l, < c_I(l,l_I): l_I \in N_I^+(l) >&=Decision_a^l(s_a(t_a + dt_a)),\\
		\forall a \in \mbox{TA}_l, <c_I(l,l_I): l_I \in N_I^+(l) >&=Decision_a^l(p_a(t^l)),
	\end{align}
with $L_I = \{ l \in L :  \forall l_I \in N_I^+(l), t^l \leq t^{l_I}  \vee t^l + dt^l <  t^{l_I} + dt^{l_I} \}$.

The set of temporary influences in a level $l \in L$ at $t^l$ is defined as:
\begin{equation}
\gamma^l{'}(t^l) = \{\gamma^l(t^l)   \bigcup_{l_I \in N_I^-(l)} c_I(l_I,l) \}.
\end{equation}

Finally, reactions are computed for levels that meet the condition defined in eq.~\ref{coherenceR}:
\\~\\
$\forall l \in L_R,$
 \begin{equation}
	\delta^l( t^l +dt^l) = Reaction^l(\sigma^l(t^l),  \gamma^l{}'(t^l)),
\end{equation}
with $L_R = \{ l \in L :  t^l +dt^l  \in min(t + dt) \}$.

The algorithm~\ref{complexalgo} summarizes this simulation model. 

\begin{algorithm}[H]
%\SetLine
\KwIn{$<L, E_I, E_P >, A(t_0), \delta(t_0)$}
\KwOut{$\delta(T)$}
\ForEach{$l \in L$}{
   $t^l = t_0$\;
}
\While{$\exists t^l \leq T$}{
   \ForEach{$a \in A(t)$}{
      $L_P = \{ l \in L : a \in A_l(t) \wedge \forall l_P \in N_P^+(l),  t^l \geq t^{l_P}\}$\;
      $p_a(t^l) = <Perception_a^l(<\delta^{l_P}(t^{l_P}): l_P \in N_P^+(l) >) : l \in L_P>$\;
      \If{$a \in \mbox{HA(t)}$}{
        $s_a(t_a + dt_a) = Memorization_a(p_a(t^l))$\;  
      }
   }
  $L_I = \{ l \in L : \forall l_I \in N_I^+(l), t^l \leq t^{l_I}  \vee t^l + dt^l <  t^{l_I} + dt^{l_I} \}$\;
   \ForEach{$l \in L_I$}{
              $<c_I(l,l_I) : l_I \in N_I^+(l) > = Natural_\omega^l(\delta^l(t^l))$ \;
              \ForEach{$a \in \mbox{HA}_l(t)$}{
                    $< c_I(l,l_I): l_I \in N_I^+(l) > = Decision_a^l(s_a(t_a + dt_a))$\;
               }
               \ForEach{$a \in \mbox{TA}_l(t)$}{
                    $<c_I(l,l_I): l_I \in N_I^+(l) > = Decision_a^l(p_a(t^l))$\;
                }
   }
   $L_R = \{ l \in L : t^l +dt^l  \in min(t + dt) \}$\;
   \ForEach{$l \in L_R$}{
          $\gamma^l{'}(t^l) = \{\gamma^l(t^l)   \bigcup_{l_I \in N_I^-(l)} c_I(l_I,l) \}$\;
          $\delta^l( t^l +dt^l) = Reaction^l(\sigma^l(t^l),  \gamma^l{}'(t^l))$\;
          $t^l = t^l +dt^l$\;
   }
}
\caption{simulation model of IRM4MLS with level-dependent temporal dynamics}
\label{complexalgo}
\end{algorithm}

\section{Discussion, conclusion and perspectives}

In this paper,  a meta-model of ML-ABM, called IRM4MLS, is introduced. It is designed to handle many situations encountered in ML-ABM: hierarchical or non-hierarchical multi-level systems with different spatial and temporal dynamics, multi-level agents or environments and agents that are dynamically introduced in levels. Moreover, IRM4MLS relies on a general simulation model contrary to the existing works published in literature. While this model is, in general, complicated, its implementation could be simplified to be more efficient in specific situations (single perception function, reactive simulation, etc.). Afterwards, examples of typical ML-ABM situations as well as ideas to treat them in the context of IRM4MLS are presented. 

In some models an agent can belong to different levels:
\begin{itemize}
\item in the model of bio-inspired automated guided vehicle (AGV) systems presented in~\citet{Morvan:2009a},  an AGV (a micro level agent) can become a conflict solver  (a macro level agent) if a dead lock is detected in the system,
\item in the SIMPOP3 multi-level model an agent representing a city  plays the role of interface between two models and then is member of two levels~\citep{Pumain:2009}. 
\end{itemize}
The simulation of these models has been addressed using different strategies:
\begin{itemize}
\item in the first example (a control problem), a top-first approach is used: the higher level takes precedence over the lower one,
\item in the second example (a simulation problem), levels are executed  alternately.
\end{itemize}
These solutions are context-dependent and likely to generate bias. In IRM4MLS, the multi-level agent situation is handled by a single simulation model that generalizes the two previous ones without scheduling bias, thanks to the influence/reaction principle.

In many multi-level agent-based models, interactions between entities in a level affect the population of agents in another level. \textit{E.g.}, in RIVAGE, a model of runoff dynamics, macro level agents (representing water ponds or ravines) emerge from micro level agents (representing water balls) when conditions are met~\citep{Servat:1998}. Then, the quantity and the flow of water become properties of macro level agents: water balls are no longer considered as agents. Conversely, micro level agents can emerge from macro level agents. Similar situations can be found in hybrid modeling of traffic flows~\citep{El-hmam:2006}. In IRM4MLS, the introduction of an agent $a$ in a level $l$ is performed by the reaction function of $l$  that introduces environmental properties representing the physical state of $a$ in $\sigma^l(t)$. Conversely, the reaction function can delete an agent from the level. An agent that does not belong to any level is inactive but can be reactivated later.

Finally, the definition of IRM4MLS is not closed in order to offer different possibilities of implementation or extension. \textit{E.g.}, levels could be defined \textit{a priori} or discovered during the simulation~\citep{Gil-Quijano:2009}. While this approach has never been used in any model so far, it seems particularly promising. In IRM4MLS, only the first possibility has been handled so far. It would be necessary to consider $L$ and $<L, E_I>$ and $<L, E_P>$ as dynamic directed graphs.

The two main perspectives of this work are the design of a modeling and simulation language and a platform that comply to the specifications of IRM4MLS as well as the re-implementation of existing models to demonstrate the capabilities of the meta-model and its simulation models.

\section*{Acknowledgments}

Authors would like to thank Javier Gil-Quijano (LIP6 -- Université Paris VI, France), Fabien Michel (LIRMM -- Université Montpellier 2, France), Daniel Jolly (LGI2A -- Université d'Artois, France) and Luce Desmidt (HEI -- France) for their help and support.

\bibliographystyle{apalike}
\bibliography{../../Biblio}

\end{document}